# Tracing app technology: An ethical review in the COVID-19 era and directions for post-COVID-19

Saleh Afroogh[1], Amir Esmalian [2]*, Ali Mostafavi [3], Ali Akbari [4], Kambiz Rasoulkhani [5], Shahriar Esmaeili [6], Ehsan Hajiramezanali [7]

[1]  Department of Philosophy, The State University of New York at Albany, Albany, NY 12203, USA; safroogh@albany.edu

[2]  UrbanResilience.AI Lab, Zachry Department of Civil and Environmental Engineering, Texas A&M University, College Station, TX 77840, USA, amiresmalian@tamu.edu

[3]  UrbanResilience.AI Lab, Zachry Department of Civil and Environmental Engineering, Texas A&M University, College Station, TX 77840, USA; amostafavi@civil.tamu.edu

[4]  Department of Biomedical Engineering, Texas A&M University, College Station, TX 77840, USA; aliakbari@tamu.edu

[5]  Hazen and Sawyer, Irvine, CA 92620, USA. E-mail: krasoulkhani@hazenandsawyer.com

[6]  Department of Physics and Astronomy, Texas A&M University, College Station, TX 77843, US, shahriar110@tamu.edu

[7]  Department of Electrical and Computer Engineering, Texas A&M University, College Station, TX. (Ehsan Hajiramezanali is currently AI Researcher at AstraZeneca.), ehsanr@tamu.edu

*  Correspondence: amiresmalian@tamu.edu

## Abstract

We conducted a systematic literature review on the ethical considerations of the use of contact tracing app technology, which was extensively implemented during the COVID-19 pandemic. The rapid and extensive use of this technology during the COVID-19 pandemic, while benefiting the public well-being by providing information about people's mobility and movements to control the spread of the virus, raised several ethical concerns for the post-COVID-19 era. To investigate these concerns for the post-pandemic situation and provide direction for future events, we analyzed the current ethical frameworks, research, and case studies about the ethical usage of tracing app technology. The results suggest there are seven essential ethical considerations—privacy, security, acceptability, government surveillance, transparency, justice, and voluntariness—in the ethical use of contact tracing technology. In this paper, we explain and discuss these considerations and how they are needed for the ethical usage of this technology. The findings also highlight the importance of developing integrated guidelines and frameworks for implementation of such technology in the post- COVID-19 world.

# I

# Introduction

In the early months of 2020, COVID-19 started spreading around the world, and a pandemic was declared by the World Health Organization (WHO) on 11 March 2020 (Chen et al. 2021; Machida et al. 2021; Miller and Smith 2021; SCASSA 2021). To track the spread of the virus, technologies, including wearable physiological sensors (Natarajan et al. 2020) and digital contact tracing apps (CTA), were leveraged by different groups in more than 100 countries (Gupta et al. 2021; Machida et al. 2021; Mbwogge 2021; Thomas 2021). CTA refers to the technology of identifying, assessing, and managing persons who may have come into contact with an infected person to control the spread of the virus by these potential virus carriers (Hoffman et al. 2020). Digital contact tracing via CTA automates the tracing process by leveraging information gathered from sensors, such as GPS and/or Bluetooth, embedded in smartphones and other devices. Experimental evidence shows the potential usefulness of CTA during pandemics (Kawakami et al. 2021; Menges et al. 2021; Rodríguez et al. 2021); however, Menges et al. (2021) discuss significant knowledge gaps regarding the design and the implementation phase of CTA. It is still considered an unproven technology. (Menges et al. 2021) Despite the positive impact of this technology on controlling the spread of the virus and enhancing the healthcare system (Menges et al. 2021), there are critical ethical considerations that affect the usefulness, reliability, and acceptability of this technology. In this paper, we review studies that discuss ethical concerns associated with CTAs during the pandemic and provide suggestions regarding future directions for the post-COVID-19 era.

Contact tracing apps rely on sensitive private data, such as users' location and their interactions with other people (Chan et al. 2020; Cho et al. 2020b). Storing and analyzing such data introduces serious privacy and security issues. There is a tradeoff, which begs ethical scrutiny, between privacy exposure issues and the benefit of CTA to healthcare. (Fahey and Hino 2020) Furthermore, the interaction between the government and people on how the implementation of the technology could affect citizens' rights is critical issue which requires further examination (Basu 2020b; Gasser et al. 2020a). Voluntariness, a choice made without coercion, in the use of CTA is a factor to be carefully examined. For example, impaired voluntariness has been shown to lead to high anxiety levels (Klar and Lanzerath 2020; Morley et al. 2020). The lack of transparency also causes distrust and failure in further development and implementation of any beneficial AI-driven technology in healthcare system. Acceptability is another consideration that is closely related to privacy, security, transparency, and other ethical concerns. High acceptability and public participation rate are essential for the successful implementation of CTA (Abuhammad et al. 2020a). Ranisch et al. (2020) explored the benefits, risks, and limitations of CTA that clearly can play a crucial role in preventing squandering of trust and misconceptions. Basu (2020b) argued that the demonstration of trust through an emphasis on transparency is an essential consideration to instilling adequate confidence in individual users.

These apparent ethical issues are mentioned and discussed in a variety of studies, reports, and case studies. The current body of knowledge, however, lacks a systematic review of the ethical considerations of implementing CTA and a discussion of the relationships and possible resolutions of these considerations. Therefore, in this study, we conducted a systematic literature review to 1) reveal the critical ethical considerations in implementing CTA and its costs and benefits, 2) discuss the unique nature of these ethical issues, and 3) provide solutions for the transition to post COVID-19 era.

In addition, these considerations are necessary for the sustainable development of the tracing app technology. Distrust has been recognized as a significant barrier to implementation of AI systems such as CTA. Public trust could greatly impact the future development and usage of CTA (Menges et al. 2021; Oldeweme et al. 2021; Siau and Wang 2018); hence, these non-technical considerations are vital for advancement of the technical development of CTA. Despite the level of success of CTA in controlling outbreaks and reducing the spread of the virus, without attention to these ethical considerations, the long-term costs associated with wide usage of this technology can significantly outweigh its benefits. Therefore, technology builders, governments, end-users, and any parties involved in building, managing, and using CTA need to take these considerations into account to maximize the effectiveness of the technology while minimizing its short- and long-term negative implications.

The remainder of the paper is organized as follows; section II describes the methodology of the systematic review of the ethical consideration in implementing CTA. Section III presents the findings and results related to the various ethical considerations and how these considerations should be given enough attention. Section IV analytically discusses each of the empirical ethical considerations and moral issues to elicit the main causes of the concerns as well as the sub-issues, which should be addressed; concluding Remarks and Future Directions for the post-COVID-19 era are also discussed at the end of this section.

# II

# Methodology

We conducted an inclusive and systematic review of the academic papers, reports, case studies, and ethical frameworks written in English. Given that there is not a specific database on the ethics of COVID-19 in general or ethics of tracing apps technology in particular, we used the Preferred Reporting Items for Systematic Reviews and Meta-Analyses (PRISMA) framework to develop a protocol in this review (figure 1).

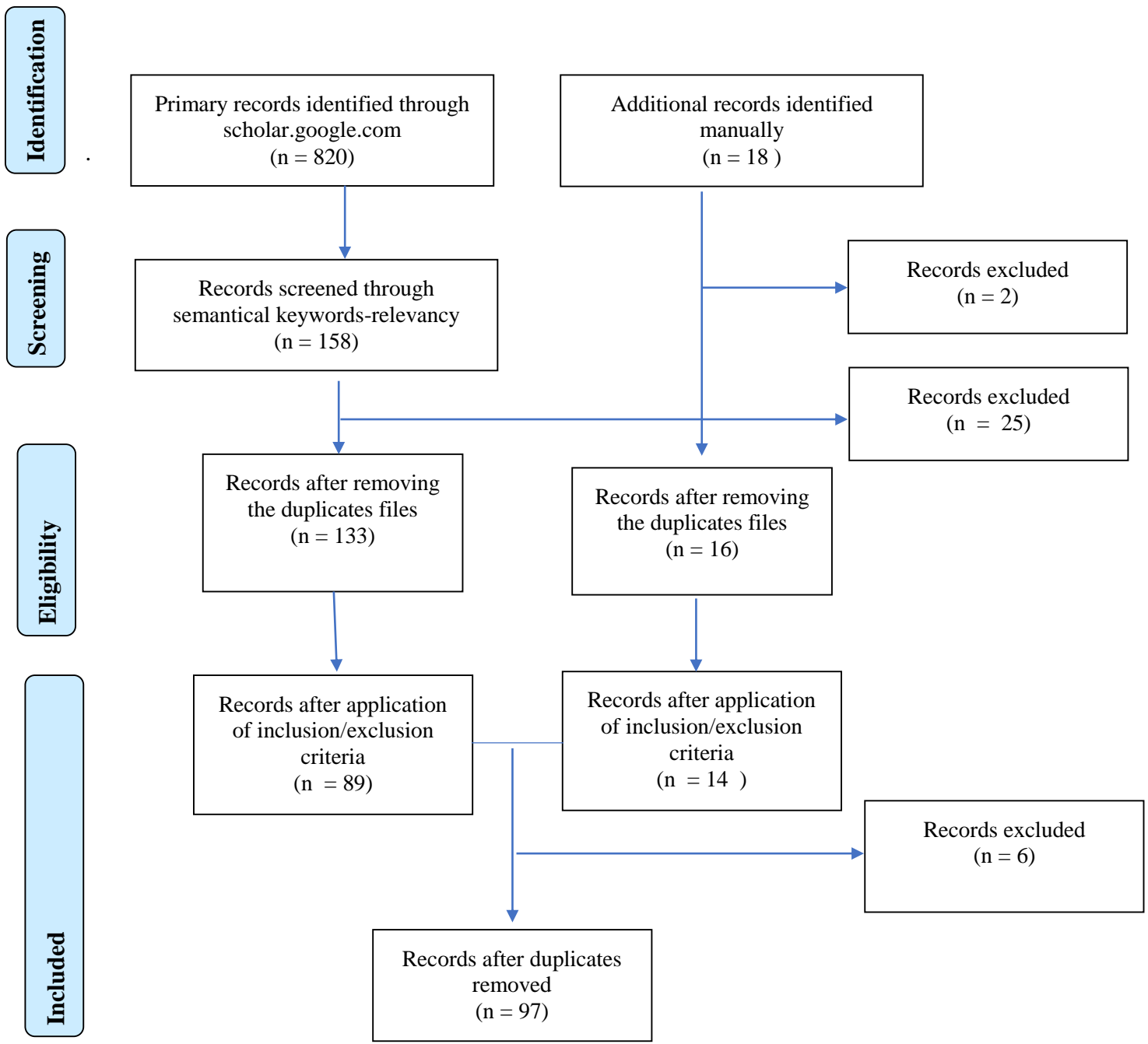

**Fig**. Developed PRISMA flow diagram for ethical review of tracing apps technology

In order to conduct a comprehensive review of the relevant studies, we followed two approaches. First, we manually searched for the most related papers on the ethical considerations of CTA: 16 papers were identified through the online search after removal of duplicate files. Secondly, we fulfilled a keyword-based search (using the http://scholar.google.com search engine) to collect all relevant papers on the topic. This search was accomplished using the following keyword phrases: (1) "tracing and COVID 19 ethics," which provided 19 relevant result pages of Google Scholar, (2) "tracing + COVID-19 + morality," for which the first five result pages were reviewed, (3) "privacy + COVID-19 + ethics," for which the first 15 result pages were reviewed; and (4) "digital surveillance + COVID-19 + ethics," for which the first 13 result pages of Google Scholar were reviewed. The last two keywords, i.e. (3) and (4), were included because of their central role in the research as two main known (based on a preliminary review) ethical considerations of tracing apps technology. Also, the search was suspended within results for each search term due to limited appearances of new relevant papers on the following pages.

The results of the search were 158 relevant papers (which were selected based on the semantical keywords relevancy), out of 820 (which appeared on the result pages). Afterward, the duplicated papers were eliminated from the analysis; we selected the 97 target papers for this systematic review based on the following two inclusion/exclusion criteria. First, articles that were published in academic journals were included. Second, the dominant topic of the papers (or a significant part of it) was the ethics of CTA. To this end, the papers' main sections were reviewed to understand their dominant topic rather than only relying on the title and papers' keywords.

Finally, the qualitative analysis on the 97 papers was performed by three researchers who critically read the papers and who developed the eight major key codes as the building blocks of the categorization of the review result in the next step of this research (Table 1).

**Table 1**
Major and minor codes included in the reviewed papers

| Major ethical codes | Number of reviewed papers | Minor ethical codes |
|---|---|---|
| Privacy | 33 | Privacy concerns, personal information, anonymity, privacy-impact tradeoff, sharing, privacy from snoopers, privacy from authorities, private and public actors, long-term privacy, decentralized |
| Security | 8 | Security, protection, data loss, unauthorized access, encryption, decentralized, data sharing, anonymized data, third party access, hack |
| Government Surveillance | 16 | Government surveillance, surveillance creep, government, civil rights, surveillance |
| Acceptability | 20 | Acceptability, public trust, voluntariness, privacy, beneficence of data, country-wise regulation |
| Transparency | 8 | Transparency, independent monitoring, reliable use, explainability, accountability, responsible data |
| Voluntariness | 10 | Voluntary, solidarity, autonomy, compulsory, anxiety, mandatory use |
| Justice | 9 | Justice, fairness, consistency, inclusion, equality, equity, (non-)bias, (non-) discrimination, diversity, plurality, accessibility, reversibility, remedy, redress, access |
| Case studies | 11 | Social liberty, transparency, ethical and legal challenges |

# III

# Findings and results

The emergence of digital public health technologies for dealing with COVID-19 calls for new and appropriate ethical frameworks fitting the unique circumstances. WHO issued Ethical Considerations to Guide the Use of Digital Proximity Tracking Technologies for COVID-19 Contact Tracing in May 2020. Several other studies (Afroogh 2021; Alanoca et al. 2021; Bruneau et al. 2020; Gasser et al. 2020a; Leslie 2020a; Morley et al. 2020; Ranisch et al. 2020; World Health Organization 2020) have proposed some ethical principles as recommendations for decision-makers in using these technologies. O'Connell et al. (2021) also proposed two kinds of *technical considerations* and *clinical and societal considerations* as practical guidelines.

The many questions surrounding ethical and practical dimensions regarding implementation of tracing apps need to be answered by decision-makers (Morley et al. 2020; Ranisch et al. 2020). There are two common value classes addressed by these questions on the ethical usage of tracing app technology. Some of these questions concern *substantive values* (which refers to the evaluative metrics of the outcomes of measures): public health benefit/effectiveness, harm minimization, privacy, justice/equity/discrimination, liberty/–autonomy/voluntariness, solidarity, surveillance (Morley et al. 2020; Ranisch et al. 2020). Other questions are pertinent to *procedural values* (which refers to guiding metrics in decision making), such as transparency, proportionality, general trustworthiness, reasonableness, accountability, consistency, engagement, reflexivity (Morley et al. 2020; Ranisch et al. 2020).

In response to these critical questions, the following major ten ethical codes to guide the decision-makers in using the tracing technologies are proposed: (i) all technologies should be temporal in design or practice (Morley et al. 2020; Ranisch et al. 2020; World Health Organization 2020), (ii) effectiveness of the technologies need to be tested before their widespread use to ensure their functionality in public health (Bruneau et al. 2020; Gasser et al. 2020a; Leslie 2020a; Morley et al. 2020; World Health Organization 2020); (iii) data collection and technology use should be limited to the minimum and "necessary amount of data" (World Health Organization 2020); (iv) any commercial use of the data must be prohibited (Leslie 2020a; World Health Organization 2020); (v) technologies should respect the individual's autonomy and have to be voluntary for all individuals to download and use the relevant apps to contribute to public health (Gasser et al. 2020a; Leslie 2020a; Morley et al. 2020; Ranisch et al. 2020; World Health Organization 2020); (vi) all the processing steps, including data collection, data retention, data storage (i.e., decentralized or centralized approach), and data analysis ought to be transparent, informed. and available for individuals (Bruneau et al. 2020; Gasser et al. 2020a; Leslie 2020a; Ranisch et al. 2020; World Health Organization 2020); (vii) high-security measures (including encryption, servers, applications, storage, etc.) must be taken seriously (Bruneau et al. 2020; Gasser et al. 2020a; Leslie 2020a; Ranisch et al. 2020; World Health Organization 2020); (viii) there should be an independent oversight body set to ensure the realization of the ethical considerations. (World Health Organization 2020); (ix) all technologies should include free participation of public health, legal, and civil society agencies (World Health Organization 2020). (x) These technologies should be justice-oriented and equity-sensitive and avoid being "digital divide" (Bruneau et al. 2020; Leslie 2020a; Ranisch et al. 2020; World Health Organization 2020); and they should not be used as a new tool to increase the government surveillance and power against citizens (Bruneau et al. 2020; Leslie 2020a; Ranisch et al. 2020; World Health Organization 2020).

While these ethical frameworks propose the general essential recommendations for implementing the tracing app technology, further analysis and explanation of ethical considerations are needed for the effective ethical use of contact tracing apps. These discussions are also needed for shedding light on the strategies to be implemented for transition to the post-COVID-19 era. Following is a discussion of these ethical considerations from an analytical perspective. These considerations are driven by the review of the papers, reports, frameworks, and case studies (Table 2).

**Table 2**
Titles, methodology, and research questions/themes of selected papers

| Title | Methodology | Research questions/themes |
|---|---|---|
| A national survey of attitudes to COVID-19 digital contact tracing in the Republic of Ireland | Survey | A national survey of the Irish population to examine barriers and levers to the use of a contact tracing application during the COVID-19 pandemic. |
| A Survey of COVID-19 contact tracing apps | Survey | Review of CTA design |
| Assessment of physiological signs associated with COVID-19 measured using wearable devices | Original empirical research | Usage of wearable devices for COVID-19 detection |
| Automated and partly automated contact tracing: a systematic review to inform the control of COVID-19 | Qualitative review paper | A systematic review of contact-tracing technologies used for the control of different diseases |
| Blind-sided by privacy? Digital contact tracing, the Apple/Google API and big tech's newfound role as global health policymakers, | Theoretical | The tole of big-tech actors in privacy |
| Contact tracing apps and values dilemmas: A privacy paradox in a neoliberal world | Theoretical, case study | The adoption of the app generates important risks to our informational privacy, surveillance, and habituation to security policies. |
| Contact tracing mobile apps for COVID-19: Privacy considerations and related tradeoffs | Theoretical, Case study | Privacy considerations and related tradeoffs, The implications of a Singaporean app during the pandemic and the ways of ameliorating the privacy concerns without decreasing usefulness to public health. |
| COVID-19 and contact tracing apps: Ethical Challenges for a social experiment on a global scale | Commentary, Theoretical, quantitative review | This paper discusses a number of questions that should be addressed when assessing the ethical challenges of mobile applications for digital contact-tracing of COVID-19, The questions addressing the ethical challenges of mobile applications used for COVID-19 contact-tracing |
| COVID-19 and the Rise of Participatory SIGINT: An examination of the rise in government surveillance through mobile applications | Original empirical research | The deployment of surveillance technologies by governments for controlling the spread of COVID-19. |
| COVID-19 Contact-tracing technology: Acceptability and ethical issues of use | Questionnaire, survey | Determine the acceptability of COVID-19 contact-tracing technology and ethical issues of use. The purpose of this study was to determine the acceptability of COVID-19 contact tracing technology and ethical issues of use. A survey study to determine the acceptability of COVID-19 contact-tracing technology and ethical issues associated with it |
| COVID-19 mobile positioning data contact tracing and patient privacy regulations: Exploratory search of global response strategies and the use of digital tools in Nigeria | Case Study; Original empirical research | The strategies for COVID-19 tracing and how using mobile positioning data conforms with Nigeria's data privacy regulations. |
| Digital contact tracing, privacy, and public health | Quantitative review | Need for transparency and accountability |
| Digital contact tracing and exposure notification: ethical guidance for trustworthy pandemic management | Theoretical, Framework, Case study | Complexity and multiplicity of the ethical considerations by presenting an ethical framework for a responsible design and implementation of CTA. Public trust is of major importance for tracing apps. |
| Digital contact tracing and surveillance during COVID-19: General and child-specific ethical issues | Review | Protect children from harm created through the collection of the data |
| Digital contact tracing, privacy, and public health | Theoretical | Privacy and public health needs in pandemic |
| Digital health innovation: exploring adoption of COVID-19 digital contact tracing apps, | Quantitative approach | Individual's intention to adopt COVID-19 digital contact tracing |
| Digital technologies in the public-health response to COVID-19 | Review paper | Capture the breadth of digital innovations for the public-health response to COVID-19 worldwide and their limitations, and barriers to their implementation, including legal, ethical and privacy barriers, as well as organizational and workforce barriers |
| Digital tools against COVID-19: Framing the ethical challenges and how to address them | Commentary, Original empirical research, Survey | Proximity and contact tracing, symptom monitoring, quarantine control, and flow modeling. For each of these, the authors discuss context-specific risks, cross-sectional issues, and ethical concerns. A typology of the primary digital public health applications currently in use and their associated risks and ethical issues. |
| Digital tools against COVID-19: taxonomy, ethical challenges, and navigation aid | Theoretical | Ethical and legal boundaries of deploying digital tools for disease surveillance, including context-specific risks, cross-sectional issues, and ethical concerns |
| Ethical considerations to guide the use of digital proximity tracking technologies for COVID-19 contact tracing: interim guidance | Normative ethical framework | The interim guidance to inform public health programs and governments that are considering whether to develop or implement digital proximity tracking technologies for COVID-19 contact tracing. |
| Ethical guidelines for covid-19 tracing apps | Ethical guidelines | Is this contact-tracing app ethically justifiable? |
| Effective contact tracing for COVID-19 Using Mobile Phones: An ethical analysis of the mandatory use of the Aarogya Setu application in India | Case Study | The mandatory requirement for installation of a COVID-19 tracing application in India and its legitimate public health intervention. Effectiveness of mobile apps in establishing a legitimate public health intervention during a public health emergency |
| Establishing prison-led contact tracing to prevent outbreaks of COVID-19 in prisons in Ireland | Case Study; Original empirical research | The process for implementation of an in-prison contact tracing technology for control of the spread of COVID-19 in Ireland. |
| Ethical framework for assessing manual and digital contact tracing for COVID-19 | Theoretical, Framework | Analyzes important technical and ethical issues of CTA + presents a framework for assessing contact tracing |
| Ethics and governance for digital disease surveillance | Theoretical | Ethical concerns raised by digital technologies and new data sources in public health surveillance during epidemics |
| Ethics and informatics in the age of COVID-19: challenges and | Theoretical - consensus-building process, Case study | Informatics-related ethical issues in light of the pandemic across three themes: (1) public health reporting and data |
| recommendations for public health organization and public policy, | | sharing, (2) contact tracing and tracking, and (3) clinical scoring tools for critical care. |
| Ethics of digital contact tracing and covid-19: Who is (not) free to go? | Case study | The practical, technical, legal and ethical considerations involved are complex, and uptake, privacy, security, and testing access have been identified as potential barriers to effectiveness. |
| Ethics of instantaneous contact tracing using mobile phone apps in the control of the COVID-19 pandemic | Theoretical, Commentary, Case study, discussion | Ethical considerations that need to be addressed in any deployment of CTA technology. Discuss ethical implications of the use of mobile phone apps in the control of the COVID-19 pandemic. Outline some ethical considerations needed for multidimensional public health response |
| Effective contact tracing for COVID-19 using mobile phones: An ethical analysis of the mandatory use of the Aarogya Setu application in India | Theoretical | Ethical analysis of mandatory use in India |
| In defence of digital contact-tracing: Human rights, south Korea and COVID-19 | Quantitative review and survey | Is the use of digital contact-tracing is valid and sufficient? |
| Information technology and the pandemic: a preliminary multinational analysis of the impact of mobile tracking technology on the COVID-19 contagion control | Review, Difference-In-differences test | Benefits and drawbacks of government surveillance |
| Integrating emerging technologies into COVID-19 contact tracing: Opportunities, challenges and pitfalls | Quantitative review and opinion | How to improve efficiency and effectiveness in the utilization of emerging technologies in contact tracing while observing the security and privacy of people in fighting the COVID-19 pandemic? Analyzing possible opportunities and challenges of integrating emerging technologies into COVID-19 contact tracing. |
| Mind the App—Considerations on the ethical risks of COVID-19 Apps | Ethical framework | Are we building the right app? |
| COVID-19 mobile positioning data contact tracing and patient privacy regulations: Exploratory search of global response strategies and the use of digital tools in Nigeria | Exploratory review | Survey strategies for digital contact tracing for the COVID-19 pandemic and to present how using mobile positioning data conforms with Nigeria's data privacy regulations. |
| PACT: Privacy sensitive protocols and mechanisms for mobile contact tracing | Theoretical, Opinion | The objective of PACT is to set forth transparent privacy and anonymity standards, |
| pandemic surveillance and racialized sub-populations: Mitigating vulnerabilities in COVID-19 app | Quantitative review and opinion | Reviewing the major challenges and a set of values that should be considered when implementing disease surveillance technology in the pandemic response |
| Preservation of privacy in the face of COVID-19: Personal data and the global pandemic | Survey | Challenges for the new models of responsible and transparent data in future public health emergencies |
| Racism and discrimination in COVID-19 responses | Quantitative review | Investigate whether the pandemic has uncovered social and political fractures within communities? |
| South Africa's COVID-19 tracing database: Risks and rewards of which doctors should be aware | Theoretical | Privacy issues associated with the deployment of CTA in Africa, especially privacy usage tradeoff. |
| Tackling COVID-19 through responsible AI innovation: Five steps in the right direction | Quantitative review and opinion, Ethical Framework & Guidelines | Investigate how better equip the data science and AI/ML community to cope with future pandemics and to support a more humane, rational, and just society? A practice-based path to responsible AI design and discovery focused on open, accountable, equitable, and democratically governed processes and products |
| The ethical imperatives of the COVID-19 pandemic: A review from data ethics | Review, Case study | Ethical imperatives observed in this pandemic from a data ethics perspective. present some ethical imperatives observed in this pandemic |
| The ethics and value of contact tracing apps: International insights and implications for Scotland's COVID-19 response | Theoretical | Privacy issues associated with the deployment of CTA in Scotland. |
| The ethics and value of contact tracing apps: International insights and implications for Scotland's COVID-19 response | Case Study | The implementation of digital contact tracing technology for pandemic management in Scotland. |
| The ethics of COVID-19 tracking apps—challenges and voluntariness | Theoretical, Ethical framework | Voluntary use. Design approach in the development of CTAs to protect of individual rights to voluntary use |
| The thorny problems of COVID-19 contact tracing apps: The need for a holistic approach | Theoretical/Opinion | What policymakers may need to consider (both technical and behavioral issues—thorny issues), especially if we are to successfully deal with the predicted second wave of COVID-19. |
| The value and ethics of using technology to contain the COVID-19 epidemic | Theoretical, Editorial | Provides an overview of the ways in which technology can support non-pharmaceutical interventions during the COVID-19 epidemic and outlines the ethical challenges associated with these approaches. The ways in which technology can support non-pharmaceutical interventions during the COVID-19 epidemic and outlines the ethical challenges associated with these approaches. What are the practical values and ethical considerations of using tracing apps technology? |
| The value and ethics of using technology to contain the COVID-19 epidemic | Survey | Ethical challenges associated with the technology can support non-pharmaceutical interventions during the COVID-19 |
| The world after coronavirus | Bulletin | Long-term consequences of our actions taken to address COVID-19 pandemic |
| Towards a seamful ethics of COVID-19 contact tracing apps? | Theoretical | Examines the development of a QR code-based tracking app as an alternative to the Bluetooth and GPS-based approaches. |
| Towards an ethics for telehealth | Survey | Ethical issues of transparency |
| Tracing surveillance and auto-regulation in Singapore: 'smart' responses to COVID-19 | Case study | The development of a Singaporean app for COVID-10 tracing and the critical debates around the use of the app. |
| Tracking the debate on COVID-19 surveillance tools | Quantitative review | Ceratin considerations should be given to approaches which are developed through machine learning type of models. |

## Privacy

Several privacy concerns are associated with the contact tracing apps since these apps continuously measure information, including users' location and their interaction with others (Mbunge 2020; Mello and Wang 2020; Scassa 2021; Subbian et al. 2020). Experimental surveys (Sowmiya et al. 2021) have revealed citizens' unwillingness to share data due to privacy concerns. Some privacy infringements of these apps, however, are justified given their potentially positive role in saving lives and reducing enormous suffering from adverse impacts of propagation of diseases (Parker et al. 2020; Sharma et al. 2020; Suh and Li 2021). Kolasa et al. (2021) verify that "contact tracing apps with high levels of compliance with standards of data privacy tend to fulfill public health interests to a limited extent. Simultaneously, digital technologies with a lower level of data privacy protection allow for the collection of more data." However, this is considered by Ishmaev et al. (2021) as a "false dilemma" and a tradeoff between privacy concerns associated with tracing apps and their positive health impacts has been an essential topic in ethical considerations among researchers (Bruneau 2020; Ekong et al. 2020b; Klar and Lanzerath 2020; Leslie 2020b; White and van Basshuysen 2021). In light of COVID-19 pandemic, the view of not sharing any private information for any reason has been given less attention since the privacy infringement is less intrusive than population-level lockdowns (Parker et al. 2020). In other words, contingent on taking necessary precautions into account when designing CTA technology, given their positive impact, these apps should be leveraged to reduce the suffering and mortality rate (Basu 2020b; Chan et al. 2020; Gasser et al. 2020b; Klaaren et al. 2020; Klar and Lanzerath 2020; Martinez-Martin et al. 2020). On the other hand, there are some external beliefs against using this technology, given their more complicated and hidden privacy issues (Osman et al. n.d.; Rowe 2020).

Privacy is a broad term that can be defined in different levels: privacy from snoopers, privacy from contact, and privacy from authorities (Cho et al. 2020b; Klar and Lanzerath 2020). The first two can more or less be addressed by appropriate design of the technology, but the privacy from authorities has been a source of concern discussed in different studies (Osman et al. 2020). Critical questions concern the power and control that different actors hold over these technologies. These questions include not only public health authorities but also other governmental agencies (e.g., police, immigration, local authorities), quasi-governmental organizations (e.g., universities), third-sector bodies (e.g., elder care services), technology companies (e.g., providers of operating systems, software, data hosting platforms), and various "shadow" players (e.g., health insurers, food retailers, credit reference agencies, data brokers) (Pagliari 2020b). Furthermore, the presence of private actors like Big Tech companies introduces additional privacy concerns, including unauthorized access to the data, dependency on corporate actors, and manipulating public policy by private actors (Henderson 2021; Sharon 2020).

In line with different levels of privacy, long-term privacy concerns are another aspect that should be considered. These issues are not obvious at first glance, and they are not limited to the pandemic period (Parker et al. 2020). The nature of access to and use of the privacy-sensitive personal data, now and in the future, necessitate accountability, transparency, and clear governance processes (Carter et al. 2020). Therefore, tracing technologies not only pose a risk to privacy, but they also put people at risk of surveillance and habituation to security policies, discrimination, and distrust, which may generate further health problems in the long term (Harari 2020).

### Security
In addition to privacy concerns and given the nature of data that can be collected by contact tracing apps on smartphones, there is a need for multiple protections against data loss and unauthorized access to the data (Sowmiya et al. 2021). One way to maintain privacy and security at the same time is to encrypt and store the data on users' phones (i.e., decentralized approach) (Dwivedi et al. 2020). This information is shared only upon request or when users test positive for the disease (Cho et al. 2020b). Storing only anonymized and aggregated data and limiting data storage to the time when a person can be contagious is another way to protect data security (Dubov and Shoptawb 2020a). Security is tightly coupled to privacy by nature; therefore, people who are worried about the security of their private data may be less willing to utilize CTA. Moreover, the tradeoff between effectiveness and security should be considered when designing CTA. Hence, data destruction protocols and use limitations, as well as reliable data security protocols preventing a third party from accessing data, are vital components of ethical framework for digital epidemiology and technological solutions such as CTA (Mbunge 2020). Furthermore, security issues might cause mental health problems such as stress, anxiety, and depression among users who have serious concerns about sharing their private and personal data (Ahmed et al. 2020). Therefore, significant attention to improving cybersecurity, especially the leveraging of government's storage and servers, is essential to alleviate such issues (Basu 2020b; Subbian et al. 2020).

**Government Surveillance**

Contact tracing systems, such as symptom checkers, are tools of syndromic surveillance that collect, analyze, interpret, and disseminate health-related data (Braithwaite et al. 2020; Gasser et al. 2020a). The COVID-19 pandemic has promoted the implementation of widespread contact-tracing techniques by governments worldwide. In fact, governments are employing surveillance technology, mainly developed through mobile-based applications, to monitor citizens, healthcare organizations, and research institutions in order to identify, locate, and track COVID-19 infected individuals and those exposed to them (Basu 2020a). In some countries, including Singapore, Israel, Italy, South Korea, Russia, Kazakhstan, and the Gulf states, the use of a COVID-19 tracing technology has been declared mandatory (Abuhammad et al. 2020a). It has been triggered by the governments as a mechanism to quickly respond by effective identification of virus infections to achieve an efficient allocation of resources to decrease the rate of infection (Gasser et al. 2020a).

Although COVID-19 surveillance technologies help governments minimize and control the spread of the outbreak, the extensive governmental investments in digital contact-tracing applications have been viewed with both cross-sectional and domain-specific ethical and legal challenges (Gasser et al. 2020a; He et al. 2021). Since the ethical and legal boundaries of implementing digital tools for COVID-19 surveillance and control purposes are unclear, the major suspicion raised by civil liberty groups is due to the perceived threat of mass surveillance (Basu 2020a).

The major concern vis-à-vis civil liberties pertains to the extension of the temporary restrictions of surveillance to a more permanent suspension of rights and liberties, which could lead to inadvertent consequences (Lucivero et al. 2020). In other words, the governments could use or abuse the surveillance technology by increasing monitoring measures even after the end of the pandemic (Abuhammad et al. 2020a). Results of a survey conducted by O'Callaghan et al. (2020) showed that 59 percent of people avoid installing a surveillance App due to a fear of the government using the App technology for greater surveillance after the pandemic. History has also proved that surveillance measures emerging during crises that were supposed to expire at a certain time are prone to be renewed or repurposed regularly (Bernard et al. 2020). As an instance, the sweeping intelligence reforms deployed by the United States under the Patriot Act following the 2001 terrorist attacks in New York granted a unique surveillance power to the government, which has never been rolled back (Bernard et al. 2020). This prospect, which is referred to as "surveillance mission creep," has been raised as a hazard that merits sustained critical attention during COVID-19 (Leslie 2020a). The possibility exists that governments could repurpose the COVID-19 surveillance technologies, which are meant to be solely used for managing the pandemic, for other governmental functions, such as policing techniques (Pagliari 2020a). It is evident that any surveillance mission creep and stealthy insertion of additional app features could erode civil liberties and privacy over time (Pagliari 2020a).

**Acceptability**
High acceptability and public participation rate are essential for the successful implementation of CTA: "a large number of individuals is required to download contact tracing apps for contact tracing to be effective" (Saw et al. 2021). Acceptability itself is tightly coupled with multiple ethical concepts and trust in CTA (Idrees et al. 2021). Privacy, voluntariness, and beneficence of the data collected by CTA are the most important metrics that affect acceptability of this technology (Abuhammad et al. 2020b; Samuel et al. 2021). Overall, when people perceive the benefits and effectiveness of this technology and its positive impact on their health, they are more willing to share their data and accept some of the privacy implications (Martinez-Martin et al. 2020). Therefore, from the perspective of ethics, it is required that all the actors, including government and private sector actors, keep the end users aware of all details and implications associated with CTA and avoid hiding or communicating false information for the sake of increasing public participation rate (Amann et al. 2021; Bradshaw et al. 2021; Fast and Schnurr 2021; Von Wyl et al. 2021) also verifies that, in the short term "the adoption of contact-tracing apps can be stimulated by monetary compensation."

Acceptability rates vary significantly between countries depending on establishment of sustained and well-founded public trust and confidence (Bunker 2020), regulations, social norms, and individuals' perceptions of costs and benefits (Chan et al. 2020; Parker et al. 2020). Studies in the United States, United Kingdom, Germany, France, and Jordan have reported support rates ranging from 42% to 80% for CTA (Abuhammad et al. 2020b; Lewandowsky et al. 2021; Lo and Sim 2020; Ranisch et al. 2020). The lack of clarity about COVID-19 tracing contact applications, including objectives, description of the application, how it works, sponsors of this technology, potential burdens to use, possible benefits, and the voluntariness for using such technology, have been found to be the most significant impediments affecting participation rate of such technology (Abuhammad et al. 2020b; Leslie 2020b). For example, even though the number of people in Jordan who agreed to the use of COVID-19 contact-tracing technology was 71.6%, the percentage of people who were using this technology was 37.8% (Abuhammad et al. 2020b). This participation rate is far less than 60% usage which has been mentioned as the required threshold for achieving maximum effectiveness (Lo and Sim 2020).

**Transparency**
Subbian et al. (2020) investigated the transparency and argued that amendments are needed for preventing COVID-19 data from being exploited. Laws and/or regulations may mandate complete transparency about what data are being used and how the data are managed and implemented in both the short and long term (Subbian et al. 2020). Also, it should be clear that using CTA is implemented on a trial basis, and its use should be subject to independent monitoring and evaluation (Lucivero et al. 2020).

Basu (2020b) argued that specific consideration is needed for building confidence in the reliable use of tracing apps. The demonstration of trust through an emphasis on transparency in data collection and its application are essential considerations for instilling adequate confidence in the reasonable individual, even in the absence of voluntariness (SCASSA 2021). Sweeney (2020) stated that implementing data-driven, machine learning-type models, in general, is very risky. Hence, approaches such as CTA applications require much higher transparency, explainability, and accountability for what data collection is currently being conducted. In this regard, the Korean Government has promoted transparency during the outbreak, publishing their findings and sending text notifications about how citizens can protect themselves. They explored the idea that a lack of transparent use of CTA apps can result in squandering public trust and raising misconceptions (Ranisch et al. 2020). Almeida et al. (2020) studied the challenges that demonstrate the need for new models of responsible and transparent data and technology governance to control SARS-CoV2 and future public health emergencies.

**Voluntariness**

Studies in ethics of implementing CTA (Abuhammad et al. 2020b; Dubov and Shoptawb 2020b; Klar and Lanzerath 2020) state that voluntariness needs to be preserved at each step of digital contact-tracing implementation—decisions to carry a smartphone, download the contact-tracing app, leave this app operating in the background, react to its alerts, and decisions to share contact logs when testing positive for COVID-19. Certain studies (Klar and Lanzerath 2020; Morley et al. 2020) take it a step further by connecting the impaired voluntariness to high anxiety levels. Voluntariness is impeded when a government threatens to impose either a lockdown or mandatory use of the tracking apps (Klar and Lanzerath 2020). Even if people are compelled to act due to a de facto social outcome in which peer pressure and expectations make using the application strongly expected, the use of CTA could be highly problematic (Morley et al. 2020). Voluntariness and consent-based data sharing is believed to be one of the most ethical approaches to mitigate the privacy risks of using tracking apps (Dubov and Shoptawb 2020b; Gasser et al. 2020b). There are, however, certain barriers in implementing such an approach to mitigate risks. Language barriers, lack of comprehension, and absence of choice are considered among the most important issues hindering implementation of consent procedures (Dubov and Shoptawb 2020a).

**Justice**

Public health emergencies actions raise important justice questions because, in these situations, infringements of justice, discrimination, and stigma commonly occur (Emanuel et al. 2020; Parker et al. 2020). CTA contributes to fairness risks over and above the general fairness risk associated with discriminatory mitigation measures. Therefore, there is a risk that CTA increases the ensuring fairness problem (Klenk and Duijf 2020). Data used in CTA may include ethnic information (race, clan, region), demographic details (gender, age, level of education, marital status), and socioeconomic status, which are subject to a variety of biases (Ntoutsi et al. 2020), can influence the allocation and distribution of COVID-19 resources for treating patients, and can ultimately lead to discrimination and riot (Mbunge 2020). Gasser et al. (2020a) have also investigated the idea that data collection should not be limited to epidemiological factors.

Gasser et al. (2020a) have emphasized that social justice and fairness should not get lost in the urgency of this crisis, and they highlight the need for meeting baseline conditions, such as lawfulness, necessity, and proportionality in AI and data processing. There is more concern about how much these technologies cost than about the injustice caused by their use. Progressively, innovation and technology will play a central role in reinforcing a dynamic plan for social justice. Hence, more attention needs to be devoted to pressing issues that exist at the nexus of technology and social justice and how social justice can address these issues most effectively. The need for researchers to act quickly and globally in tackling COVID-19 demands unprecedented practices of open research and responsible data sharing. Devakumar et al. (2020) "emphasize that health protection does not only depend on effective universal healthcare systems but relies on social inclusion, justice, and solidarity. They argue that the absence of these values leads to the escalation of inequalities, scapegoating, and long-lasting discrimination, with broad negative public (health) outcomes." Hendl et al. (2020) investigated that if apps are promoted as an integral part of the COVID-19 pandemic response, this should be done with a clear and explicit commitment to values of health equity, non-discrimination, and solidarity with vulnerable sub-populations.

**Case Studies**
As the implementation of COVID-19 infection tracing technologies has proven successful at controlling the spread of the virus in some countries, such as China and Spain (Rebollo et al. 2021), there has been growing enthusiasm for rapidly expanding such technology to other countries (Abuhammad et al. 2020a). Other countries, however, are still under unprecedented uncertainty about how to deploy these technologies to not only limit the spread of COVID-19 but also to respect their citizens' rights (Pagliari 2020a). In fact, ethical and legal challenges are presented by the implementation of contact-tracing technologies, which call for taking the extra mile to make reasonable efforts to ensure no violation of civil rights will occur. Including, but not limited to, their privacy, liberty, consent, and public benefit. In this regard, some case studies in various countries have examined the ethical and legal implications of COVID-9 tracing technologies and have discussed approaches for improving the technology concerns without inhibiting its benefits to public health.

A recent study by Cho et al. (2020) evaluated the implementation of a mobile-based COVID-19 tracing application developed in Singapore. This case study declared that privacy was a key component of considerations around this technology and offered certain modifications to be made into the existing application (such as partially anonymizing by polling or mixing different users' tokens to enhance anonymity) for stronger privacy protection (Cho et al. 2020a). A case study related to India demonstrated that the installation of a government-backed COVID-19 tracing application was mandatory in certain situations. This study argues that the mandatory application requirement represents a legitimate public health intervention during the pandemic (Basu 2020a).

In Western countries, such as the United States and the United Kingdom, the deployment of COVID-19 surveillance technologies has raised issues, such as public trust and data privacy, which necessitated some considerations in the technology design to make a balance between public benefits and pandemic risks (Pagliari 2020a). Using the contact-tracing technology in Scotland for pandemic management outlines challenges and opportunities for public engagement and raises ethical questions to make informed decisions at multiple levels, from application design to institutional governance (Pagliari 2020a). The COVID-19 surveillance has been shown to significantly improve the capacity and scope of timely outbreak response in Nigeria. Although this technology was used within the current regulation of Nigeria, the existence of guidelines seemed necessary to curb abuse of the data collected through this approach (Ekong et al. 2020a). Amann et al. (2021) provides an analytic survey of the media ecosystem's ideas regarding CTA in Germany, Austria, and Switzerland and concludes that "achieving public consensus on digital contact tracing apps currently seems unlikely. To foster public trust and acceptance, authorities thus need to develop clear and coherent communication strategies that listen to and address public concerns." Investigating the adoption state of the SwissCOVID app in Switzerland during the pandemic (von Wyl et al. 2021) also argues that "communicating the benefits of digital proximity tracing apps is crucial to promote further uptake and adherence of such apps and, ultimately, enhance their effectiveness to aid pandemic mitigation strategies."

# IV

## Discussion

Digital tracing technology has the potential to transcend the tradeoff between saving lives and livelihoods by freeing people from quarantine while containing the virus (Klenk and Duijf 2020). Digital technologies have been playing an important role in a comprehensive response to outbreaks and pandemics, complementing conventional public health measures and thereby contributing to reducing the human and economic impact of COVID-19. Technology can support non-pharmaceutical interventions during the COVID-19 epidemic (Dubov and Shoptawb 2020a). An overview of the ways in which technology can support non-pharmaceutical interventions during the COVID-19 epidemic has also been provided by Budd et al. (2020). However, several requirements exist for these interventions to be ethical and to be able to ensure public confidence during the COVID-19 pandemic. It is still too early in the COVID-19 pandemic timeline to fully quantify the added value of digital technologies to the pandemic response (Budd et al. 2020).

Some types of COVID-19 technology might lead to the employment of disproportionate profiling, policing, and criminalization of marginalized groups (Hendl et al. 2020). Furthermore, there are technical limitations, dealing with asymptomatic individuals, privacy issues, political and structural responses, ethical and legal risks, consent and voluntariness, abuse of contact tracing apps, and discrimination in using CTA (Mbunge 2020; Mbunge et al. n.d.). The ethical considerations and questions pertinent to tracing technologies date back to old and fundamental ethical considerations aimed at protecting basic human and moral values and civil rights. The specific circumstances of COVID-19, however, cause a new revision of previous ethical frameworks and sheds light on the significance of the problems. Some of these ethical problems, such as privacy, the uncontrollable increasing power of government surveillance (especially in some countries which are highly susceptible to violating individuals' privacy and human rights), could entail unforeseeable negative impacts on humans' lives in the near future. Thus, we need to critically contemplate the consequences of any decisions on aspects of citizen' lives in the short, middle, and long term. We should produce inclusive, ethical frameworks for tracing app technologies both in the design/research and practice levels. These frameworks should also include various critical considerations in order to effectively account for various ethical dimensions, as discussed below.

Ethical concerns related to privacy, security, and anonymity are among the significant barriers to the use of contact tracing apps (Andrew Tzer-Yeu Chen 1967; Elkhodr et al. 2021; Mbunge et al. 2021; Smoll et al. 2021; Urbaczewski and Lee 2020). Despite these barriers, contact tracing apps have been successfully deployed in several cases for controlling the spread of the virus (Cho et al. 2020b; Mbunge 2020). Hence, the tradeoff between usability and privacy is the most important consideration when using such apps. Technological solutions are one way of addressing certain privacy concerns. For instance, GPS tracing versus Bluetooth tracing, centralized versus decentralized data processing, restricted versus extended data usage, data encryption, and anonymization techniques can be leveraged to reduce privacy risks (Subbian et al. 2020) (Chan et al. 2020). In addition to technological considerations, governments can play an important role in limiting privacy concerns through regulatory efforts (Gasser et al. 2020b; Subbian et al. 2020; Urbaczewski and Lee 2020). Given the significant effect of privacy on acceptability of this technology by general population (Subbian et al. 2020; Zimmermann et al. 2021), the government can play an important role in convincing or mandating people to use this technology in the light of Mill's classic harm principle where the "physical or moral good" of the individual is deemed able to be superseded if necessary for preventing "harm to others." (Basu 2020b). However, it must be noted that solutions are not one-size-fits-all. In other words, certain solutions that have worked for some countries may not effectively work in other countries with different societal norms. There are significant differences in individuals' perceptions when evaluating the costs and benefits relating to privacy which may require specific strategies which account for the contextual considerations (Cho et al. 2020b; Sharma et al. 2020).

Data security is a critical aspect of CTA since the health data collected by smartphones are prone to be hacked or abused by third parties. There are varying opinions about effective ways to improve security and privacy. Some researchers believe in decentralized approaches in which the data are locally saved on the user's phone and shared upon request, while others think using a centralized data storage in which government's servers and/or encrypted databases are leveraged can improve the security, especially from snoopers and hackers (Platt et al. 2021). Although governments can claim the notion of improved data security and trust, there are certain concerns associated with governmental access to private data (Greenleaf and Kemp 2021). Private data, especially data gathered about location and personal interactions, can be used by governments for inappropriate purposes. This becomes a more significant concern if the data storage is not limited to the period of pandemic or the time when a person can be contagious. White and Van Basshuysen (2021) show that "the public at large regard centralized architectures with suspicion."

Acceptability of CTA technology has a direct and significant impact on effectiveness and success of implementing this technology. This technology will be useful at the community level when it is being used by at least 60% of the population (Lo and Sim 2020). Some countries have addressed this by mandating the usage of contact tracing apps (Cho et al. 2020b; Mbunge 2020; Parker et al. 2020). However, this solution would be less applicable in democratic countries (Basu 2020b; Mello and Wang 2020). In this case, the role of technology designers and, more importantly, that of governments becomes vital, as they can improve general trust by establishing effective, transparent, accountable, and inclusive oversight as well as transparent, auditable, and easily explained algorithms, the highest possible standards of data security; and effective protections around the ownership uses of data (Parker et al. 2020).

While contact tracing will be successful only when enough people participate, Dubov and Shoptawb (2020b) believe that even under conditions of public health emergencies, no one should be obligated to share their personal information. However, (Lucivero et al. 2020) does not totally agree with this argument and believes further studies are needed to explore 1) to what extent the responsibility of a public health matter should be placed on individuals; 2) what this means in terms of accountability (delineating who is legally responsible if something goes wrong). Furthermore, Emanuel et al. (2020) and Parker et al. (2020) believe even increasing the number of participants by providing incentives should be considered carefully on case-by-case basis, as all people might not be able to benefit equally. Klar and Lanzerath (2020) pointed out two factors as prerequisites for the success of the Rakning C-19 app (with the best penetration rate of all contact trackers in the world): 1) the guarantee that all rights will be preserved, and 2) the ensuing trust in the app and the institutions that handle the data. It is clearly specified who has access to the data and how long it will be kept, the data will not be repurposed, and mission creep will be prevented.

During the COVID-19 pandemic, governments' disaster management has been dependent upon location, health, and population data to forecast the rates of infection, decrease new infections, understand the efficiency of social distancing directives, and improve the efficiency of vaccine development. Nevertheless, the pandemic surveillance technologies are triggering a complex set of ethical and legal hazards exacerbating the increasing challenges to civil liberty, autonomy, privacy, and public trust globally. Merely asserting that an application is voluntary to install or its processes, missions, and functions are visible would not allay the existing concerns (Bernard et al. 2020). Therefore, these technologies must be subject to certain oversight and regulation that oblige the governments to use them ethically, robustly, and transparently and avoid any violation of privacy rights or establishment of a dictatorial police state after COVID-19 outbreak (Abuhammad et al. 2020a; Gasser et al. 2020a).

Analysis of recent COVID-19 surveillance technology debates, controversial programs, and emerging outcomes in comparable countries applying this strategy discloses socio-technical complexities and surprising paradoxes that necessitate further research and that reveal the need for comprehensive, adaptive, and inclusive strategies in using such technology to fight the pandemic (Abuhammad et al. 2020a). In countries, China, South Korea, Ireland, Israel, Singapore, Nigeria, and India, highly privacy-invasive COVID-19 tracing approaches have been adopted to manage the spread of the virus (Clarke et al. 2020; Lee and Lee 2020). In general, their citizens have complied with utilizing the tracing technologies, whether compulsorily or voluntarily. In contrast, considering privacy as a highly ethical issue and the desire for freedom from government surveillance is very strongly felt in Europe, where concerns have led some countries, such as Ireland and Germany, to cancel their plans and change course. Different jurisdictions within countries as well as their specific micro-cultures play a key role in balancing the benefits and risks associated with the implementation of contact-tracing technologies. Therefore, it is vital to assess the performance of actors and practices to understand the real meaning of mandating the use of tracing apps in the lives of people by deploying the contact-tracing approaches (Abuhammad et al. 2020a).

# V

# Concluding Remarks and Future Directions

Review of the research literature on the ethical use of CTA implies that certain consideration is lacking in the incorporation of the technology, and there is an urgent need for developing the metrics and strategies for enabling ethical use of the technology. It is also time to evaluate the functionality of CTA on the basis of the relevant metrics (Durrheim et al. 2021).

The emergency circumstance of the COVID-19 pandemic prevents us from developing comprehensive and inclusive frameworks to address basic and fundamental moral problems. This unique circumstance calls for some short-term practical codes to maximize ethical values in emergency conditions. Such a unique circumstance, however, could be an opportunity to discuss ethical and moral consequences of computational and digital technologies and innovative development, such as empathic and value-sensitive design (Afroogh et al. 2021; Umbrello and van de Poel 2021), to make sure technology increases the population's well-being in the long term, and will not be used against them.

Privacy and security are central concerns that can even limit public willingness towards using technologies such as CTA. One of the factors that can improve the security and privacy of this technology and also enhance public's trust towards these apps is establishing regulations for destroying the collected data after a crisis. CTA can be useful even after the pandemic for tracking down other contagious diseases; however, the costs implied by security and privacy concerns may dominate the health benefits when the pandemic is over. Therefore, it is important to give people the option of whether to share their private data and also to choose the level at which they prefer to share data. A combination of local storage of data on user's phone along with proper encryption and anonymization can guarantee the users that their private data are safe from third parties as well as from snoopers.

The current costs and benefits of CTA apps are opaque to the end-users. The general public may not be aware of the benefits of these apps and their role in a pandemic; however, privacy and security issues associated with sharing their private data are tangible and concerning. As a result, many users may underestimate the benefits and overestimate the costs, which dissuades them from supporting this technology. Communicating successful examples of the deployment of this technology in cases in which it was effective in detecting outbreak clusters and mitigating impact and containment could change society's perspective (Basu 2020b; Raman et al. 2021). Overall, making people aware of the technology's details; algorithms, specifically cybersecurity considerations; and the health impacts could engender acceptability. In this case, it is important not to misrepresent or exaggerate the benefits. Simply conveying the most truthful information to the end-users and allowing them to voluntarily accept this technology are effective approaches for motivating public participation and trust.

There is a need for an independent and trustworthy institution to dispel distrust in CTA in the future. On the one hand, explicitly making the use of tracking apps compulsory seems to be more transparent, and therefore, might increase the end-users trust in CTA, compared to an in-principle voluntary but de facto constraining approach (Lucivero et al. 2020). On the other hand, advocating compulsory adoption cannot overcome the fact that certain groups within society may not be able to access this technology, which leads to inequalities (Abuhammad et al. 2020a; Lucivero et al. 2020). To this end, an independent and trustworthy institution that can handle the data, ensure privacy, and can support those who might not have access to a smartphone or internet could increase the end users' trust and the number of volunteers.

Governments willing to implement any of the emerging COVID-19 surveillance technologies need to address their ethical and legal issues. They must put safeguards into place to avoid harm and mitigate the remaining risks. The issues raised around government surveillance remind us that any COVID-19 surveillance program needs to respect people's privacy, have transparency at its core, protect the collected data, limit surveillance to the minimum necessary to overcome the current crisis, address potential issues of discrimination, adhere to values of democracy, and clarify upfront the duration or timeline of operation. In addition, certain procedural guidance and frameworks must be established as a navigation aid in the form of an iterative set of steps to work through and meet baseline principles, such as adaptivity, flexibility, reflexivity, transparency, accountability, responsiveness.

(Lai et al. 2021) predicts the expansion of digital CTA to complement the human-based contact tracing for future pandemics, while the recent case studies have highlighted the importance of transparency, accountability, and stakeholder participation for the credibility of digital tracing strategies in controlling the pandemic. In democratic societies, there is a need for reasonable efforts to instill confidence among citizens for using these applications in order to effectively manage the pandemic. In nondemocratic societies, however, legitimate concerns exist over surveillance creep through applications, and culture of lax civil rights, where political protesters may be tracked and suppressed using such technologies after a pandemic (Basu 2020a). In this regard, civil society organizations must warn against both the mandatory use of such technologies during pandemics and data misuse by data handlers after the pandemics. Therefore, best practices yet have to emerge for COVID-19 tracing technologies.